\documentclass[11pt]{article}
\usepackage{amsmath}
\usepackage{amsfonts}
\usepackage{amssymb}
\usepackage[dvips]{epsfig}

 \advance \textwidth by -2in
 \advance \textheight by -3in
\def\##1{{\bf #1}}
\def\=#1{\underline{\underline{#1}}}

\def\+#1{\underline{\bf #1}}
\def\*#1{\underline{\underline{\bf #1}}}

\def\eps{\epsilon}

\def\.{\mbox{ \tiny{$^\bullet$} }}

\def\ux{\#{u}_x}
\def\uy{\#{u}_y}
\def\uz{\#{u}_z}

\def\le{\left(}
\def\ri{\right)}
\def\les{\left[}
\def\ris{\right]}
\def\lec{\left\{}
\def\ric{\right\}}

\def\r#1{(\ref{#1})}

\begin{document}

\begin{center} {\bf {\LARGE Electrical control of the
linear optical properties of particulate composite materials}} \end{center} \vskip 0.2cm

\noindent  {\bf Akhlesh Lakhtakia$^a$ and Tom G. Mackay$^b$}
\vskip 0.2cm

\noindent {\sf $^a$ CATMAS~---~Computational \& Theoretical Materials Sciences Group \\
\noindent Department of Engineering Science \& Mechanics\\
\noindent 212 Earth \& Engineering Sciences Building\\
\noindent Pennsylvania State University, University Park, PA
16802--6812, USA\\
email: akhlesh@psu.edu} \vskip 0.4cm

\noindent {\sf $^b$ School of Mathematics\\
\noindent James Clerk Maxwell Building\\
\noindent University of Edinburgh\\
\noindent Edinburgh EH9 3JZ, United Kingdom\\
email: T.Mackay@ed.ac.uk} \vskip 0.4cm

\begin{center} {\bf Abstract} \end{center}

The Bruggeman formalism for the homogenization of particulate
composite materials is used to predict the effective permittivity
dyadic of a two--constituent composite material with one constituent
having the ability to display the Pockels effect. Scenarios wherein
the constituent particles are randomly oriented, oriented spheres,
and oriented spheroids are numerically explored. Thereby,
homogenized composite materials (HCMs) are envisaged whose
constitutive parameters may be continuously varied through the
application of a low--frequency (dc) electric field. The greatest
degree of control over the HCM constitutive parameters is achievable
when the constituents comprise oriented and highly aspherical
particles and have high electro--optic coefficients.


\vskip 0.4cm

\section{Introduction}
In 1806, after having ascended the skies in a balloon to collect
samples of air at different heights and after having ascertained the
proportions of different gases in each sample, Jean--Baptiste Biot
and Fran\c{c}ois Arago published the first known homogenization
formula for the refractive index of a mixture of mutually inert
gases as the weighted sum of their individual refractive indexes,
the weights being in ratios of their volumetric proportions in the
mixture (Biot \& Arago 1806). The Arago--Biot mixture formula
heralded the science and technology of particulate composite
materials~---~particularly in optics, more generally in
electromagnetics, and even more generally in  many other branches of
physics. An intensive literature has developed over the last two
centuries in optics (Neelakanta 1995; Lakhtakia 1996), and recent
forays into the realms of metamaterials and complex mediums
(Grimmeiss {\em et al.} 2002; Weiglhofer \& Lakhtakia 2003; Mackay
2005) have reaffirmed the continued attraction of both particulate composite
materials and homogenization formalisms.

Post--fabrication dynamic control of the effective properties of a mixture of two constituent materials is
a technologically important capability underlying the successful deployment of a
host of smart materials and
structures. Dynamic control can be achieved in many ways,
particularly if controlability and sensing capablity are viewed
as complementary attributes.  One way is to infiltrate the composite
material with another substance, possibly a fluid, to change, say, the effective optical response
properties (Lakhtakia {\em et al.} 2001; M\"onch {\em et al.} 2006). This can be adequate
if rapidity of change is not  a critical requirement. Another way is
to tune the effective properties by the application of pressure
(Finkelmann {\em et al.} 2001; Wang {\em et al.} 2003) or change of temperature (Schadt \& F\"unfschilling 1990).
Faster ways of dynamic control could involve the use of electric fields if one constituent material
is a liquid crystal
(Yu {\em
et al.} 2005) or magnetic fields if one constituent material is magnetic (Shafarman {\em
et al.} 1986).

Our focus in this paper is the control of the effective permittivity tensor of a
two--constituent composite material, wherein both constituent materials are classified
as dielectric materials in the optical regime but one can display the Pockels effect (Boyd 1992).
Both constituent materials can be distributed as ellipsoidal particles whose orientations can
be either fixed  or be completely random.

The plan of this paper is as follows: Section \ref{the} contains a
description of the particulate composite material of interest, as well as the
key equations of the Bruggeman homogenization formalism (Bruggeman
1935; Weiglhofer {\it et al.} 1997) adopted to estimate the relative
permittivity dyadic of the homogenized composite material (HCM).
Section \ref{nrd} presents a few numerical examples to show that the
Pockels effect can be exploited to dynamically  control the linear
optical response properties of composite materials through a
low--frequency electric field. Given the vast parameter space
underlying the Pockels effect, we emphasize that the examples
presented are merely illustrative. Some brief concluding remarks are
provided in Section~\ref{conc}.

A note about notation:  Vectors are in boldface, dyadics are double underlined.
A Cartesian coordinate system with unit vectors $\#u_{x,y,z}$ is
adopted. The identity  dyadic is written as $\=I$, and the null dyadic as $\=0$.
An $\exp(-i\omega t)$ time--dependence is implicit
 with $i = \sqrt{-1}$, $\omega$ as angular frequency, and $t$ as time.

\section{Theory}\label{the}

Let the two constituent materials of the particulate composite material
be labeled $a$ and $b$. Their respective
volumetric proportions are denoted by $f_a$ and $f_b=1-f_a$. They are distributed
as ellipsoidal particles. The dyadic
\begin{equation}
\=U^{(a)} = \sum_{K=1}^3\, \alpha_K\,\#a_K\#a_K
\end{equation}
describes the shape of particles made of material $a$, with
$\alpha_K>0\,\forall K \in\les1,3\ris$ and the three unit vectors
$\#a_{1,2,3}$ being mutually orthogonal. The shape dyadic
\begin{equation}
\=U^{(b)} = \sum_{K=1}^3\, \beta_K\,\#b_K\#b_K
\end{equation}
similarly describes the shape of the particles made of material $b$.
A low--frequency
(or dc) electric field $\#E^{dc}$ acts on the composite material, the prediction of whose effective
permittivity dyadic in the optical regime is of interest.

Material $a$
does not display the Pockels effect and, for simplicity, we take it to be
isotropic with relative permittivity scalar $\eps^{(a)}$ in the
optical regime.

Material $b$ has  more
complicated dielectric properties as it displays the Pockels effect. Its
linear electro--optic
properties are expressed through the inverse
of
its relative permittivity dyadic in the optical regime, which  is written  as (Boyd 1992)
\begin{eqnarray}
\nonumber
\les\=\eps^{(b)}\ris^{-1}&=&
\sum_{K=1}^3\les\le1/\epsilon _{K}^{(b)}+s_j\ri \,\#u_K\#u_K\ris\\[5pt]
&+&s_4 \le\#u_2\#u_3 +\#u_3\#u_2\ri
+s_5 \le\#u_1\#u_3 +\#u_3\#u_1\ri
+s_6 \le\#u_1\#u_2 +\#u_2\#u_1\ri\,,
  \label{Poc}
\end{eqnarray}
where
\begin{equation}
\label{sJ}
s_J=
\sum_{K=1}^3 r_{JK}\,E_{K}^{dc}\,,\quad J\in\les1,6\ris\,,
\end{equation}
and the unit vectors
\begin{equation}
\left.\begin{array}{l}
\#u_1=-(\ux\cos\phi_b+\uy\sin\phi_b)
\cos\theta_b+\uz\sin\theta_b\\[5pt]
\#u_2=\ux\sin\phi_b-\uy\cos\phi_b\\[5pt]
\#u_3=(\ux\cos\phi_b+\uy\sin\phi_b)\sin\theta_b+\uz\cos\theta_b
\end{array}\right\}\,,\quad \theta_b\in\les0,\pi\ris\,,\quad
\phi_b\in\les0,2\pi\ris\,,
\end{equation}
are
relevant to the crystallographic structure of the material. In
\r{Poc} and \r{sJ},
$E_{K}^{dc}=\#u_K\.\#E^{dc}$, $K\in\les1,3\ris$, are the Cartesian components of the dc
electric field; $\epsilon _{1,2,3}^{(b)}$ are the principal relative
permittivity scalars in the optical regime; and $r_{JK}$, $J\in\les1,6\ris$
and $K\in\les1,3\ris$, are the $18$ electro--optic coefficients in the
traditional contracted or abbreviated notation for representing symmetric
second--order tensors (Auld 1990).  Correct to the first order in the
components of the dc electric field, which is commonplace in electro--optics (Yariv \& Yeh 2007), we
get the linear approximation (Lakhtakia 2006a)
\begin{eqnarray}
\nonumber
\=\eps^{(b)}
&\approx& \sum_{K=1}^3\les\epsilon _{K}^{(b)}\le
1-\epsilon _{K}^{(b)}s_K \ri\,\#u_K\#u_K\ris\\[5pt]
&-&\epsilon _{2}^{(b)}\epsilon _{3}^{(b)}\,s_4 \le\#u_2\#u_3 +\#u_3\#u_2\ri
-\epsilon _{1}^{(b)}\epsilon _{3}^{(b)}\,s_5 \le\#u_1\#u_3 +\#u_3\#u_1\ri
-\epsilon _{1}^{(b)}\epsilon _{2}^{(b)}\,s_6 \le\#u_1\#u_2 +\#u_2\#u_1\ri
\label{PocEps}
\end{eqnarray}
from (\ref{Poc}), provided that
\begin{equation}
\label{restriction}
\lec
{\substack{\max \\ K\in\les1,3\ris}} \,\lvert \eps_K^{(b)}\rvert\ric\,\,
\lec{\substack{\max \\ J\in\les1,6\ris}}\,\lvert s_J\rvert \ric\ll 1\,.
\end{equation}
This material can be isotropic, uniaxial, or biaxial, depending on the relative
values of
$\epsilon_1^{(b)}$, $\epsilon_2^{(b)}$, and $\epsilon_3^{(b)}$. Furthermore,
this material may belong to one of  20 crystallographic classes of point group
symmetry, in accordance with the relative
values of the electro--optic coefficients.

Let the Bruggeman estimate of the relative permittivity dyadic of HCM be
denoted by $\=\eps^{Br}$. If the particles of material $a$   are all
identically oriented with respect to their crystallographic axes,
and likewise the particles of material $b$, then $\=\eps^{Br}$ is
determined by solving the following equation (Weiglhofer {\em et
al.} 1997):
\begin{eqnarray}
\nonumber
&&f_a \le \eps^{(a)}\=I-\=\eps^{Br}\ri\.
\les \=I +\=d^{(a)}\.\le \eps^{(a)}\=I-\=\eps^{Br}\ri\ris^{-1}\\[5pt]
&&\qquad\quad + f_b \le\=\eps^{(b)}-\=\eps^{Br}\ri\. \les
\=I +\=d^{(b)}\.\le \=\eps^{(b)}-\=\eps^{Br}\ri\ris^{-1} =\=0\,.
\label{Brug}
\end{eqnarray}
In this equation,
$
\=d^a\equiv\=d(\=U^{(a)})$ and
$\=d^b\equiv\=d(\=U^{(b)})$,
where the dyadic function
\begin{equation}
\=d(\=U) =\frac{1}{4\pi }
\int_{\phi_q=0}^{2\pi}
\int_{\theta_q=0}^{\pi}
\sin\theta_q\,
\frac{\=U^{-1}\.\#q\#q\.\=U^{-1}}
{\#q\.\=U^{-1}\.\=\eps^{Br}\.\=U^{-1}\.\#q}\,
d\theta_q\,d\phi_q\,
\end{equation}
contains the unit vector
\begin{equation}
\#q=\uz\cos\theta_q + (\ux\cos\phi_q+\uy\sin\phi_q)\,\sin\theta_q\,.
\end{equation}

The Bruggeman formalism is more complicated when the relative permittivity
dyadics of the
particles of material   $b$   are randomly oriented. To begin with,
let there be $P\ge 1$ distinct orientations.
We represent the p$^{th}$ orientation of the relative permittivity
dyadic $\=\eps^{(b)}$  in terms of  the set of Euler angles
$\lec\gamma_{n_p}\ric_{n=1}^{3}$ as (Lakhtakia 1993)
\begin{equation}
\=\eps^{(b)}_p=
\=R_{\,z}(\gamma_{3_p})\.\=R_{\,y}(\gamma_{2_p})\.\=R_{\,z}(\gamma_{1_p})\.
\=\eps^{(b)}\.\=R_{\,z}^{-1}(\gamma_{1_p})\.\=R_{\,y}^{-1}(\gamma_{2_p})\.\=R_{\,z}^{-1}(\gamma_{3_p})\,,\qquad
p = 1,2,3,.....,P,
\end{equation}
wherein the rotational dyadics
\begin{equation}
\left. \begin{array}{l} \=R_{\,y} (\gamma)= \le \#u_x\, \#u_x +
\#u_z\, \#u_z \ri \cos \gamma + \le \#u_z\, \#u_x -  \#u_x\, \#u_z
\ri \sin \gamma + \#u_y\, \#u_y \vspace{2mm}
 \\
\=R_{\,z} (\gamma) = \le \#u_x\, \#u_x + \#u_y\, \#u_y \ri \cos
\gamma + \le \#u_x\, \#u_y -  \#u_y\, \#u_x \ri \sin \gamma +
\#u_z\, \#u_z
 \end{array} \right\}.
\end{equation}
Let us  define
\begin{equation}
 \=\Pi (\gamma_{1_p}, \gamma_{2_p}, \gamma_{3_p})= \le\=\eps^{(b)}_p
-\=\eps^{Br}\ri\. \les \=I +\=d^{(b)}\.\le \=\eps^{(b)}_p
-\=\eps^{Br}\ri\ris^{-1} .\end{equation}
 Then the Bruggeman equation
may be expressed in the form
\begin{equation}
f_a \le \eps^{(a)}\=I-\=\eps^{Br}\ri\. \les \=I +\=d^{(a)}\.\le
\eps^{(a)}\=I-\=\eps^{Br}\ri\ris^{-1} + f_b
\frac{1}{P}\,\sum_{p=1}^{P}\,\=\Pi^{} (\gamma_{1_p}, \gamma_{2_p},
\gamma_{3_p}) =\=0\,, \label{BrugP}
\end{equation}
if all $P$  orientations are equiprobable. In the limit $P
\rightarrow \infty$, equation \r{BrugP} becomes
\begin{eqnarray}
\nonumber &&f_a \le \eps^{(a)}\=I-\=\eps^{Br}\ri\.
\les \=I +\=d^{(a)}\.\le \eps^{(a)}\=I-\=\eps^{Br}\ri\ris^{-1}\\[5pt]
&&\qquad\quad +
 \frac{f_b}{8 \pi^2} \int^{2
\pi}_{\gamma_3 = 0} \int^{ \pi}_{\gamma_2 = 0} \int^{2
\pi}_{\gamma_3 = 0}  \,\=\Pi (\gamma_{1}, \gamma_{2}, \gamma_{3}) \,
\sin \gamma_2 \; d \gamma_1 \, d \gamma_2\, d \gamma_3 =\=0\,.
\label{BrugP2}
\end{eqnarray}
Even more complicated orientational averages~---~e.g., of particulate shapes and
geometric orientation of particles, in addition to crystallographic orientation~---~can
be similarly handled.

The HCM relative permittivity $\=\eps^{Br}$ can be extracted from
equation \r{Brug} and equation \r{BrugP2}  iteratively using
standard techniques, and  a Jacobi iteration technique is
recommended (Michel 2000).

\section{Numerical results and discussion}\label{nrd}

A vast parameter space is covered by the homogenization formalism described
in the previous section. The parameters include: the volumetric proportions
and the shape dyadics of materials $a$ and $b$; the relative permittivity
scalar $\eps^{(a)}$; the three relative permittivity scalars $\eps_{K}^{(b)}$
and the upto 18 distinct electro--optic coefficients $r_{JK}$ of material $b$;
the angles $\theta_b$ and $\phi_b$ that describe
the crystallographic orientation of material $b$ with respect
to the laboratory coordinate system; and
the magnitude and direction of $\#E^{dc}$.
To provide illustrative results here,
 we set
$\eps^{(a)}=1$.  All calculations were made for two choices
of material $b$ (Cook 1996):
\begin{itemize}
\item[I.] zinc telluride, which belongs
to the cubic ${\bar 4}3m$
crystallographic class:  $\epsilon_1^{(b)} = \epsilon_2^{(b)}= \epsilon_3^{(b)}=8.94$, $r_{41}=r_{52}=r_{63}=4.04
\times 10^{-12}$~m~V$^{-1}$, and all other $r_{JK}\equiv0$; and
\item[II.] potassium niobate, which belongs
to the orthorhombic $mm2$
crystallographic class: $\epsilon_1^{(b)} = 4.72$,
$\epsilon_2^{(b)}= 5.20$, $\epsilon_3^{(b)}=5.43$,
$r_{13}=34\times 10^{-12}$~m~V$^{-1}$,
$r_{23}=6\times 10^{-12}$~m~V$^{-1}$,
$r_{33}=63.4\times 10^{-12}$~m~V$^{-1}$,
$r_{42}=450\times 10^{-12}$~m~V$^{-1}$,
$r_{51}=120\times 10^{-12}$~m~V$^{-1}$, and
all other $r_{JK}\equiv0$.
\end{itemize}
Given the huge parameter space still left, we chose to fix
$f_b=0.5$, the Bruggeman formalism then being maximally
distinguished from other homogenization formalisms such as the
Maxwell Garnett (Weiglhofer {\em et al.} 1997) and the
Bragg--Pippard formalisms (Bragg \& Pippard 1953; Sherwin \&
Lakhtakia 2002). Finally, we chose particles of material $a$ and $b$
to be spherical (i.e., $\=U^{(a)} = \=U^{(b)}=\=I$) in
Sections~\ref{random} and \ref{oriented_spheres}, and spheroidal in
Section~\ref{oriented_spheroids}.

Two different scenarios based on the orientation of material $b$
were investigated. The scenario wherein the
 material $b$ particles are randomly oriented with respect to their
  crystallographic
 axes was considered in the  study presented in Section~\ref{random}.
 Particles of
 material $b$ were taken to have the same orientation with respect to their
  crystallographic
 axes in the  studies presented in Sections~\ref{oriented_spheres} and
\ref{oriented_spheroids}.

For all scenarios, the estimated permittivity dyadic of the HCM may be compactly
represented as
\begin{equation}
\=\eps^{Br}= \alpha^{Br}\, \=I + \beta^{Br} \,(\#u_M\#u_N+\#u_N\#u_M),
\end{equation}
wherein the unit vectors $\#u_M$ and $\#u_N$ are aligned with the
optic ray axes
 of the HCM (Chen 1983; Weiglhofer \& Lakhtakia 1999; Mackay \&  Weiglhofer 2001).
For the real--symmetric relative permittivity dyadic $\=\eps^{Br}$,
with three distinct
(and orthonormalised) eigenvectors
$\#e_{1,2,3}$ and corresponding eigenvalues $\epsilon_{1,2,3}^{Br}$,  the scalars
$\alpha^{Br}$ and $\beta^{Br}$ are given by
\begin{equation}
\left.
\begin{array}{l}
\alpha^{Br} = \epsilon_2^{Br}\\[5pt]
\beta^{Br} = \displaystyle{\frac{\epsilon_3^{Br} - \epsilon_1^{Br} }{2}}
\end{array}
\right\},
\end{equation}
whereas the unit vectors $\#u_{M,N}$ may be stated  as
\begin{equation}
\left.
\begin{array}{l}
\#u_M = \displaystyle{ \le \frac{\epsilon_2^{Br} - \epsilon_1^{Br}
}{\epsilon_3^{Br} - \epsilon_1^{Br} } \ri^{1/2} \#e_{1} +\le
\frac{\epsilon_3^{Br} - \epsilon_2^{Br} }{\epsilon_3^{Br} -
\epsilon_1^{Br} } \ri^{1/2} \#e_{3}   }\vspace{2mm} \\
\#u_N = \displaystyle{- \le \frac{\epsilon_2^{Br} - \epsilon_1^{Br}
}{\epsilon_3^{Br} - \epsilon_1^{Br} } \ri^{1/2} \#e_{1}+\le
\frac{\epsilon_3^{Br} - \epsilon_2^{Br} }{\epsilon_3^{Br} -
\epsilon_1^{Br} } \ri^{1/2} \#e_{3}   }
\end{array}
\right\},
\end{equation}
for $\epsilon_1^{Br} < \epsilon_2^{Br} < \epsilon_3^{Br}$.

In accordance with mineralogical literature (Klein \& Hurlbut 1985),
we  define the linear birefringence
\begin{equation}
\delta_n=\le\epsilon_3^{Br}\ri^{1/2}-
\le\epsilon_1^{Br}\ri^{1/2}\,,
\end{equation}
 the degree of biaxiality
 \begin{equation}
\delta_{bi}=
      \le\epsilon_3^{Br}\ri^{1/2}+\le\epsilon_1^{Br}\ri^{1/2}-2
      \le\epsilon_2^{Br}\ri^{1/2}\,,
      \end{equation}
      and the
angles
\begin{equation}
\left. \begin{array}{l} \delta = \displaystyle{ \cos^{-1}\les \le
\frac{\epsilon_3^{Br} - \epsilon_2^{Br} }{\epsilon_3^{Br} -
\epsilon_1^{Br} } \ri^{1/2}\ris} \vspace{2mm}\\
\theta_M = \cos^{-1} \#u_M\.\#u_z \vspace{2mm}
\\
\theta_N = \cos^{-1} \#u_N\.\#u_z  \end{array}\right\}.
\end{equation}
The linear birefringence $\delta_n$ is the difference between the   largest and the smallest refractive indexes of the HCM;
the degree of biaxiality $\delta_{bi}$ can be either positive or
negative, depending on the numerical value of
$\le\epsilon_2^{Br}\ri^{1/2}$  with respect to the mean of
$\le\epsilon_1^{Br}\ri^{1/2}$ and $\le\epsilon_3^{Br}\ri^{1/2}$; $2
\delta$ is the angle between the two optic ray axes; and
$\theta_{M,N}$ are the angles between the optic ray axes and the
Cartesian $z$ axis. Thus, $\=\eps^{Br}$ can be specified by six
real--valued parameters: $\eps^{Br}_2$, $\delta_n$, $\delta_{bi}$,
$\delta$, and $\theta_{M,N}$, in a physically illuminating
way.

\subsection{Randomly oriented spherical electro--optic particles}
\label{random}

 We begin by considering the scenario
wherein the particles of
 material $b$ are randomly oriented with respect to their
  crystallographic
 axes, and the particles of both materials are spherical.
Accordingly, the HCM is an isotropic dielectric medium,
characterized by the relative permittivity dyadic $\=\eps^{Br}=\eps^{Br}\=I$.

\begin{figure}[!ht]
\centering \psfull \epsfig{file=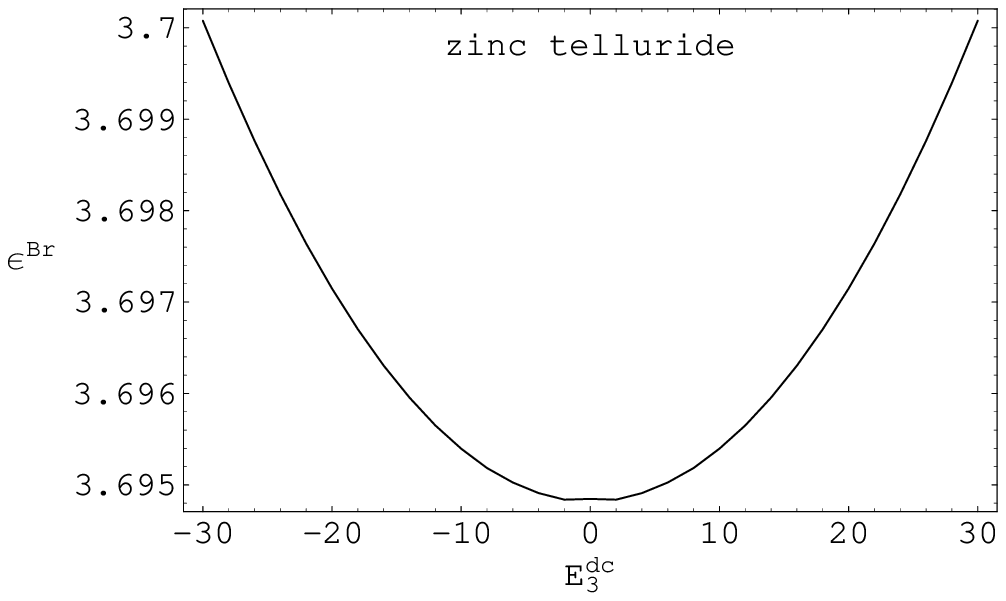,width=5.8in}
\vspace{5mm} \\  \epsfig{file=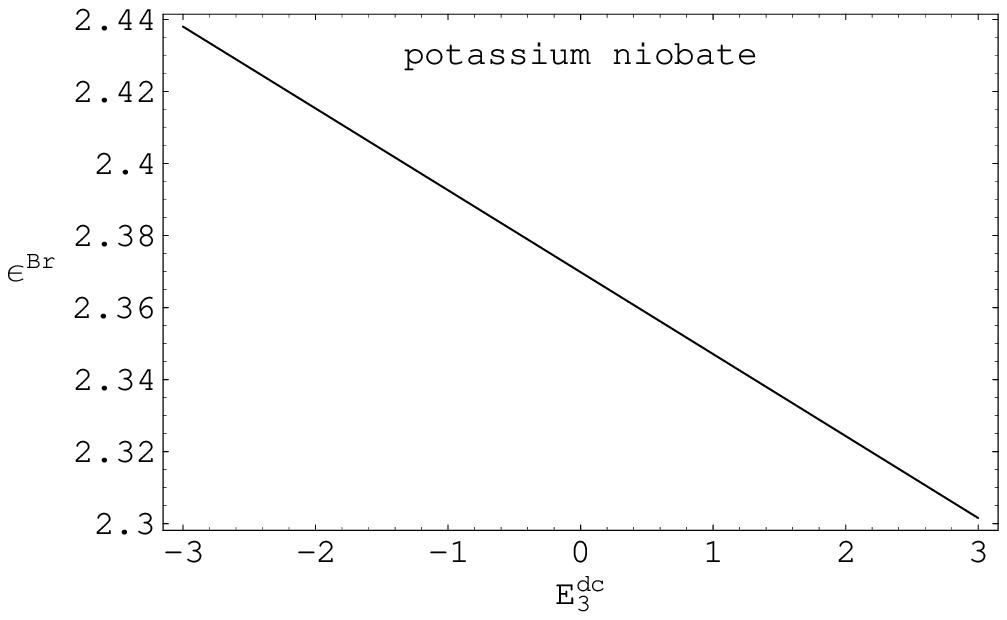,width=5.8in}
  \caption{\label{fig1}
  The estimated relative permittivity scalar $\eps^{Br}$ of the HCM
   plotted against $E^{dc}_{3}$ (in V m${}^{-1}$ $\times
  10^8$) for $E^{dc}_{1,2} = 0$ and $f_a = 0.5$. Material $b$ is zinc telluride for the upper graph and
  potassium niobate for the lower graph. The particles of material $b$  are randomly oriented with respect to
their
  crystallographic
 axes, and both types of particles are spherical.
  }
\end{figure}

The Bruggeman estimate  $\eps^{Br}$, as extracted from equation
\r{BrugP2}, is plotted in Figure~\ref{fig1} against $E^{dc}_3$, with
$E^{dc}_{1,2} =0$.  Material $b$ is zinc telluride for the upper
graph and  potassium niobate for the lower in this figure. The range for the
magnitude of $\#E^{dc}$ in Figure~\ref{fig1}~---~and for all
subsequent figures~---~was chosen in order to comply with
\r{restriction}. In the case where material $b$ is zinc telluride,
$\eps^{Br}$  varies only slightly as $E^{dc}_{3}$ changes, and
$\eps^{Br}$ is insensitive to the sign of $\eps^{Br}$. A greater
degree of sensitivity to $E^{dc}_3$ is observed for  the HCM which
arises when material $b$ is potassium niobate; in this case the
HCM's relative permittivity  is also sensitive to the sign of $E^{dc}_3$,
thereby underscoring the significance of crystallographic class
of the electro--optic constituent material even when averaging
over crystallographic orientation is physically valid.

\subsection{Identically oriented spherical electro--optic particles}
\label{oriented_spheres}

The scenario wherein all  particles of material $b$ are taken to be
identically oriented, with particles of both constituent materials
being spherical, is now considered. Let us begin
with the case where material $b$ is zinc telluride. This is an
isotropic material when $\#E^{dc}=\#0$, but the anisotropy
underlying the Pockels effect becomes evident on the application of
the low--frequency electric field (Lakhtakia 2006b).

\begin{figure}[!ht]
\centering \psfull\epsfig{file=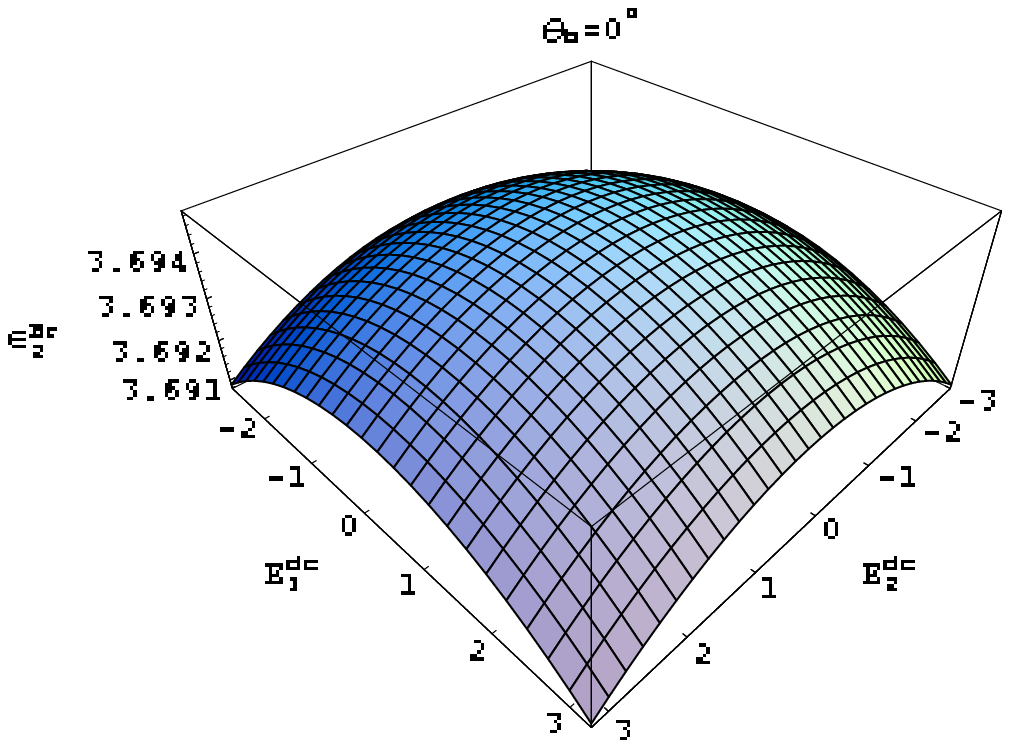,width=2.8in}
\hspace{10mm} \epsfig{file= 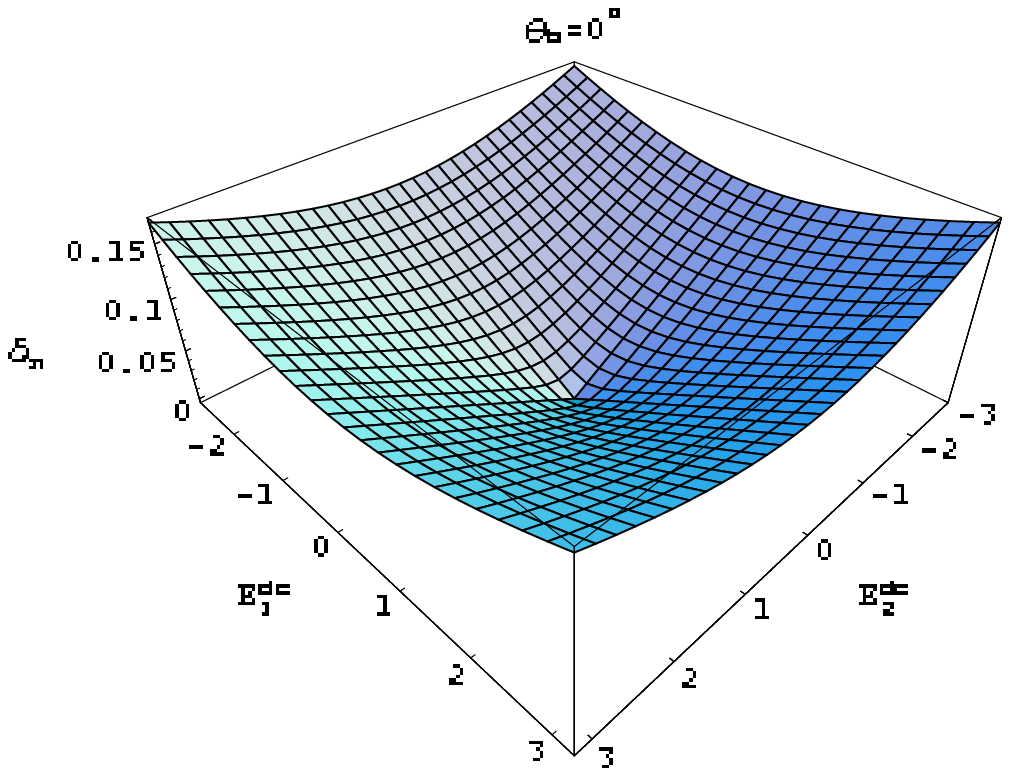,width=2.8in}\\
\vspace{10mm} \psfull\epsfig{file=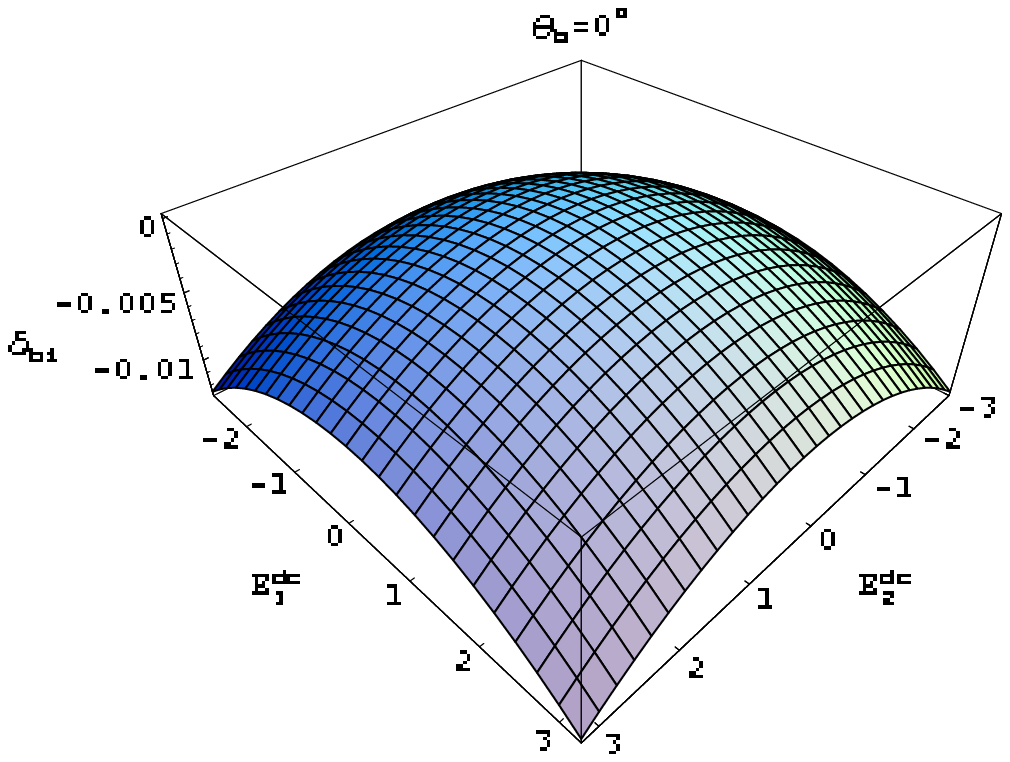,width=2.8in}
\hspace{10mm} \epsfig{file= 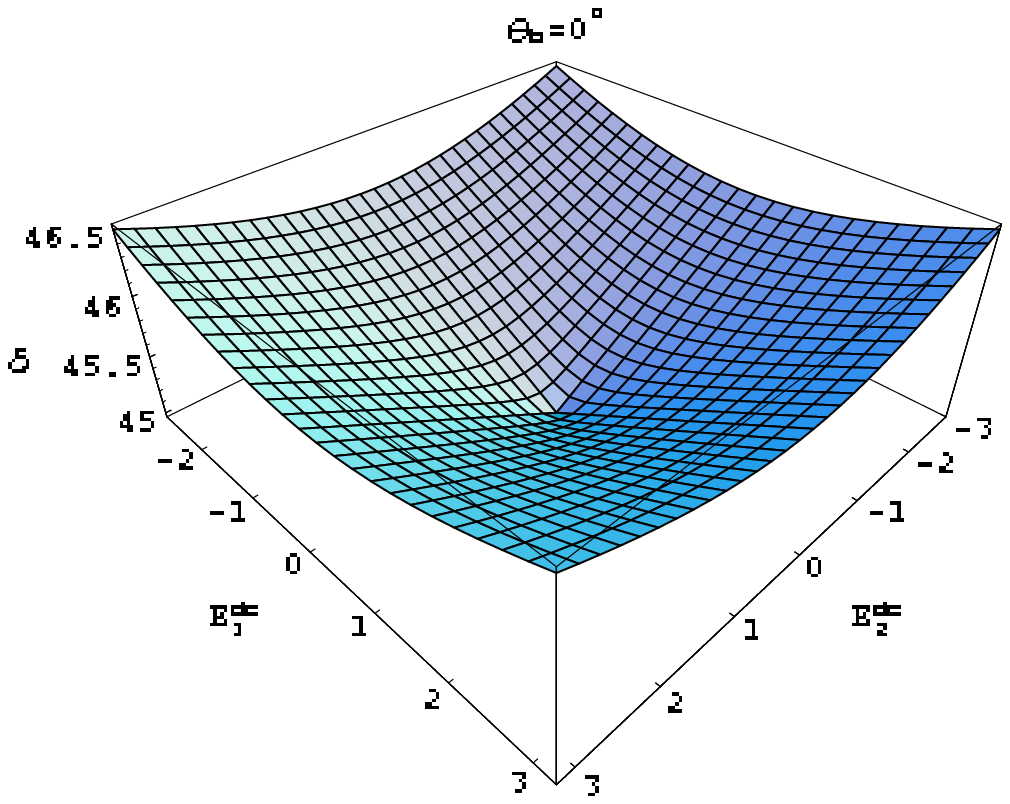,width=2.8in}
\\ \vspace{10mm}
\psfull\epsfig{file=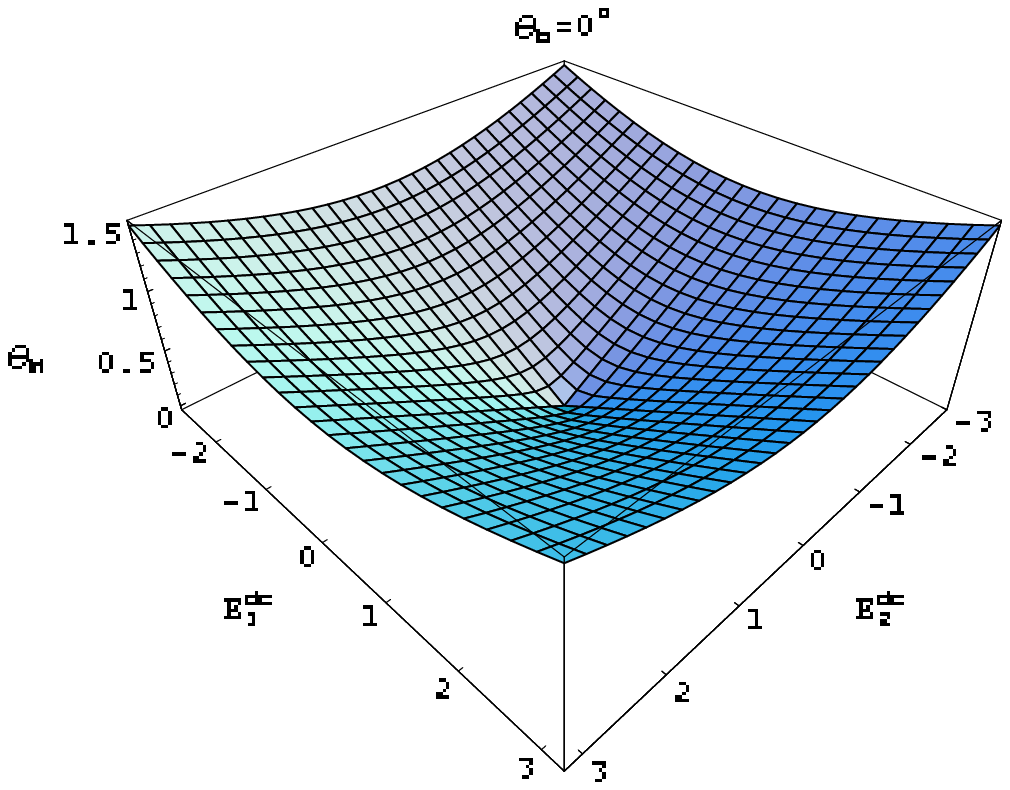,width=2.8in} \hspace{10mm}
\epsfig{file= 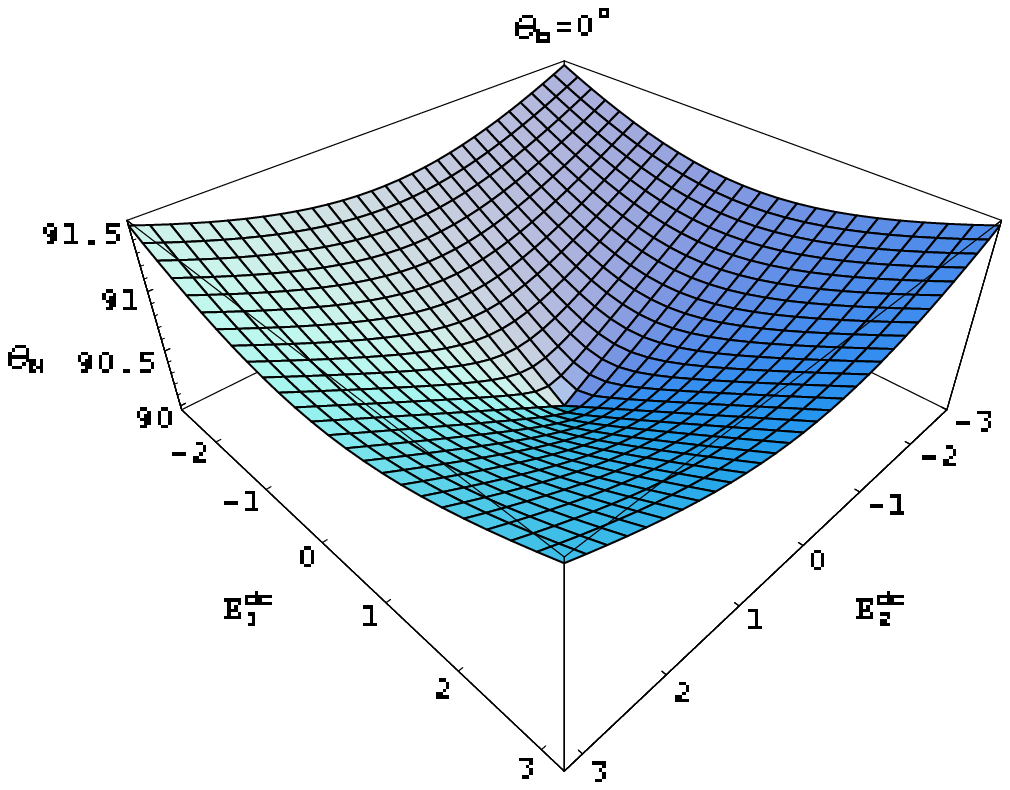,width=2.8in}
  \caption{\label{fig2} The HCM parameters $\eps^{Br}_2$,
  $\delta_n$, $\delta_{bi}$, $\delta$ (in degree) and $\theta_{M,N}$
  (in degree) plotted against $E^{dc}_{1,2}$ (in V m${}^{-1}$ $\times
  10^9$). The crystallographic orientation angles
  of material $b$ are $\theta_b
= \phi_b = 0$; and $E^{dc}_3 = 0$. Material $b$ is zinc telluride, and the
particles of both constituent materials are spherical.}
\end{figure}

 The HCM parameters, as extracted from  equation \r{Brug}, are plotted in
Figure~\ref{fig2} as functions of $E^{dc}_{1,2}$  with $E^{dc}_3
=0$. The crystallographic orientation angles $\theta_b = \phi_b = 0$.  As expected,
in this figure $\delta_n= \delta_{bi} =0$ (i.e., the HCM is
isotropic) when $E^{dc}_1=E^{dc}_2=0$. The HCM constitutive
parameters $\eps^{Br}_2$, $\delta_n$, $\delta_{bi}$, $\delta$, and
$\theta_{M,N}$ are all insensitive to the signs of $E^{dc}_{1}$ and
$E^{dc}_{2}$. The HCM is negatively biaxial in general (because
$\delta_{bi}<0$), although the biaxiality is small. The linear
birefringence is not sensitive to the signs
of $  E_{1,2}^{dc}  $; it increases considerably as $\vert \#E^{dc}\vert$
is increased, as a glance at data on minerals readily confirms
(Griblle \& Hall 1992). The two
optic ray axes remain almost mutually orthogonal, as indicated by $\delta
\approx 45^\circ$, as $\vert E_{1,2}^{dc}\vert$ are changed.

\begin{figure}[!ht]
\centering \psfull\epsfig{file=xS2_ZT_A_eps2.eps,width=2.8in}
\hspace{10mm} \epsfig{file= 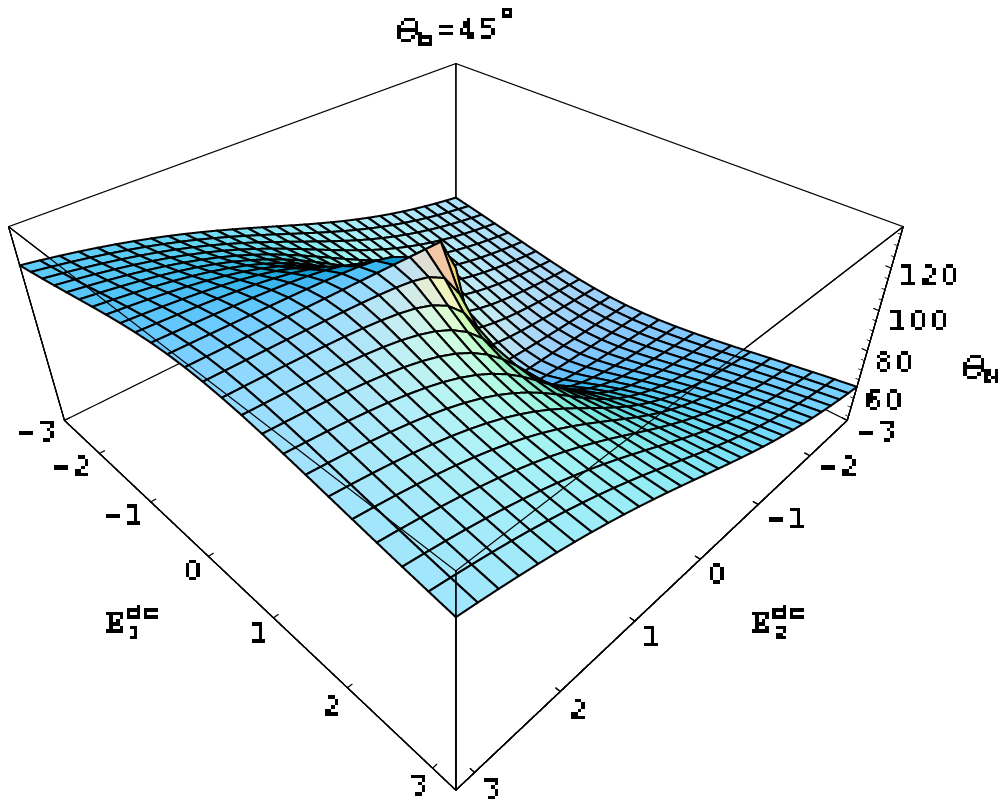,width=2.8in}
\\ \vspace{10mm}
\psfull\epsfig{file=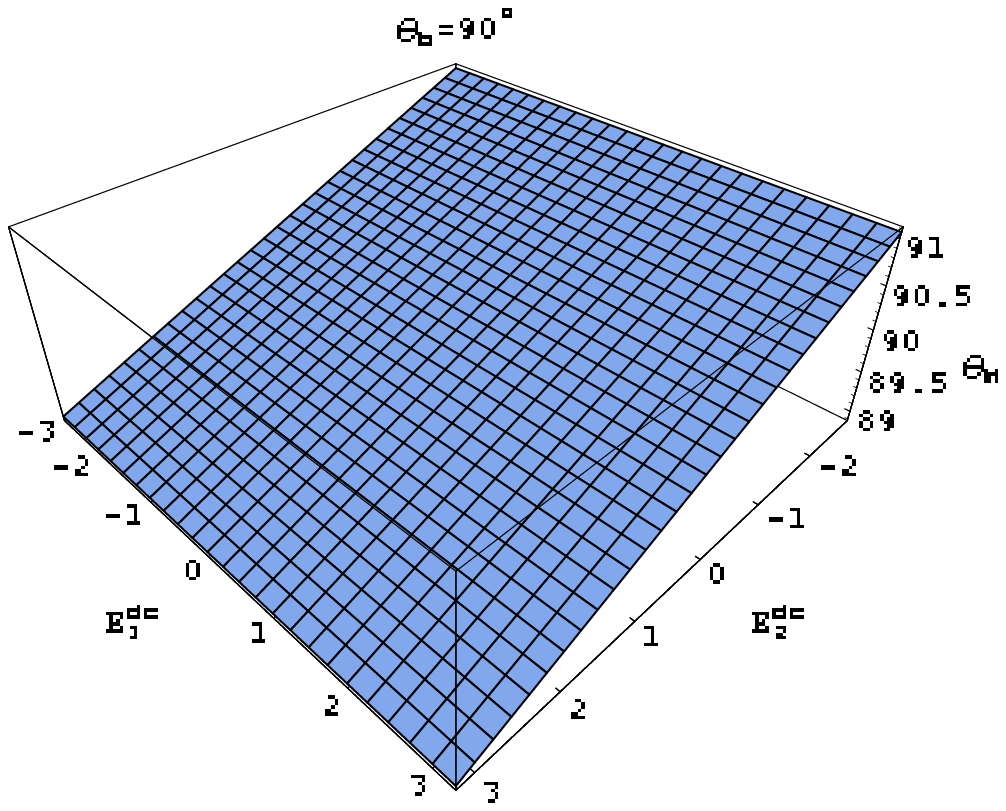,width=2.8in} \hspace{10mm}
\epsfig{file= 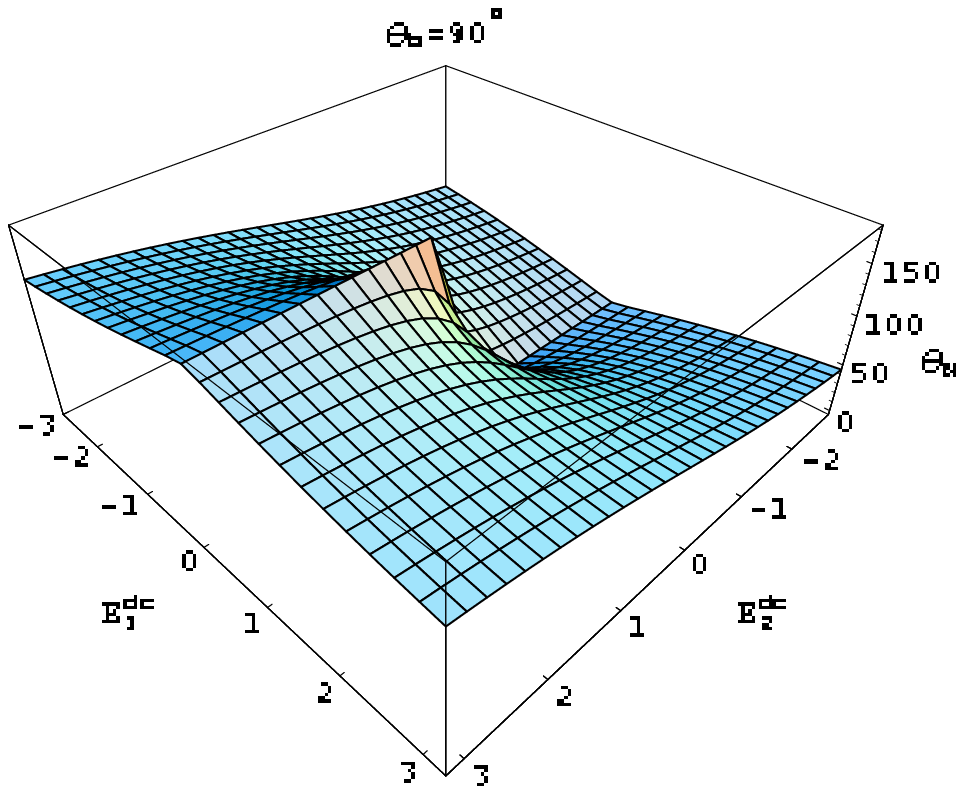,width=2.8in}
  \caption{\label{fig3}
The optic ray axis angles $\theta_{M,N}$  (in degree) of the HCM plotted
against $E^{dc}_{1,2}$ (in V m${}^{-1}$ $\times
  10^9$). The constitutive parameters
  of material $b$ are same as in Figure~\ref{fig2}
except that $\theta_b \in \lec  45^\circ, 90^\circ \ric$. (The
corresponding plots of $\eps^{Br}_2$,
  $\delta_n$, $\delta_{bi}$,
  and $\delta$ are not noticeably different to  those presented in
Figure~\ref{fig2}.)}
\end{figure}

The influence of the  orientation angle $\theta_b$ for zinc
telluride is explored in Figure~\ref{fig3}. Here, the optic ray axis
angles $\theta_{M,N}$ are plotted as functions of $E^{dc}_{1,2}$
with $E^{dc}_3 =0$, for $\theta_b \in \lec 45^\circ, 90^\circ \ric$
with $\phi_b = 0$. The orientations of both optic ray axes
continuously vary with increasing $E^{dc}_{1,2}$  in a manner which
continuously varies as $\theta_b$ increases.
 The polar angle of the optic ray axis aligned with $\#u_M$, namely
$\theta_M$, is slightly sensitive to  $E^{dc}_2$ but insensitive  to
$E^{dc}_1$.
 In
contrast, $\theta_N$ is acutely sensitive to both $E^{dc}_1$ and
$E^{dc}_2$. Furthermore, $\theta_N$ is sensitive to the sign of
$E^{dc}_2$ but not the sign of $E^{dc}_1$.  The HCM parameters
$\eps^{Br}_2$, $\delta_n$, $\delta_{bi}$,
$\delta$~---~which are not presented in Figure~\ref{fig3}~---~are
insensitive to increasing $\theta_b$; the plots for these quantities
are not noticeably different from the corresponding plots presented
in Figure~\ref{fig2}.

Let us turn now to the case where material $b$ is potassium niobate.
This material is anisotropic (orthorhombic and negatively biaxial) even
when the Pockels effect is not invoked, and it has much
higher electro--optic coefficients than zinc telluride~---~hence, it can be expected
to lead a different palette of HCM properties.

\begin{figure}[!ht]
\centering \psfull\epsfig{file=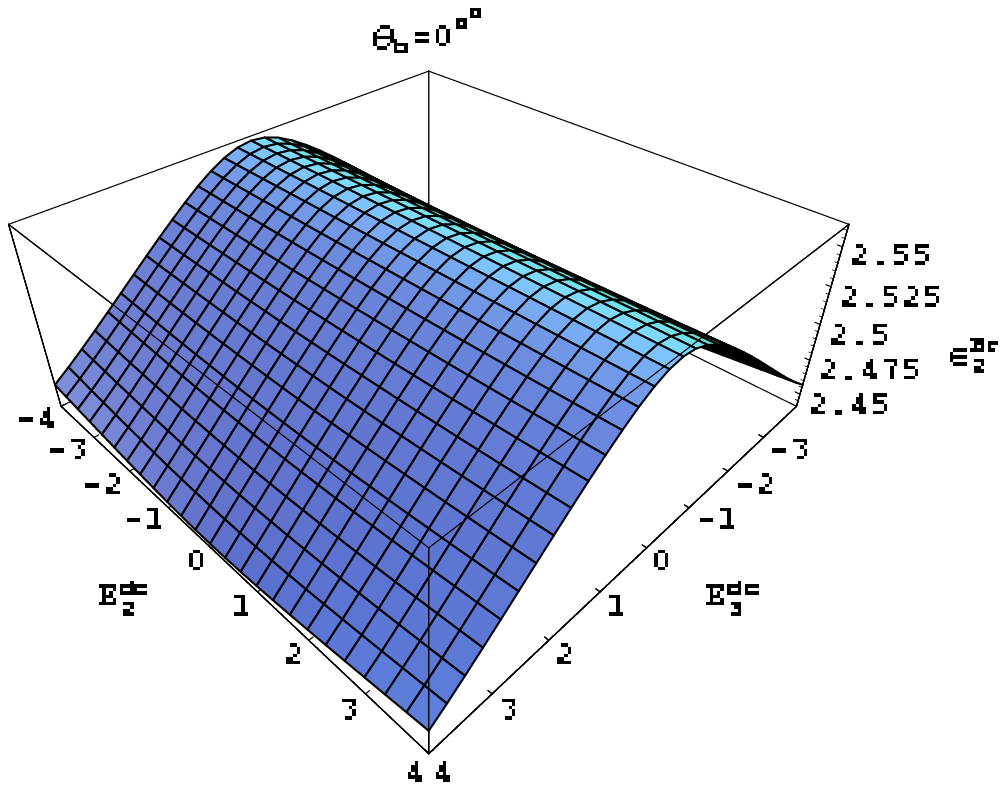,width=2.8in}
\hspace{10mm} \epsfig{file= 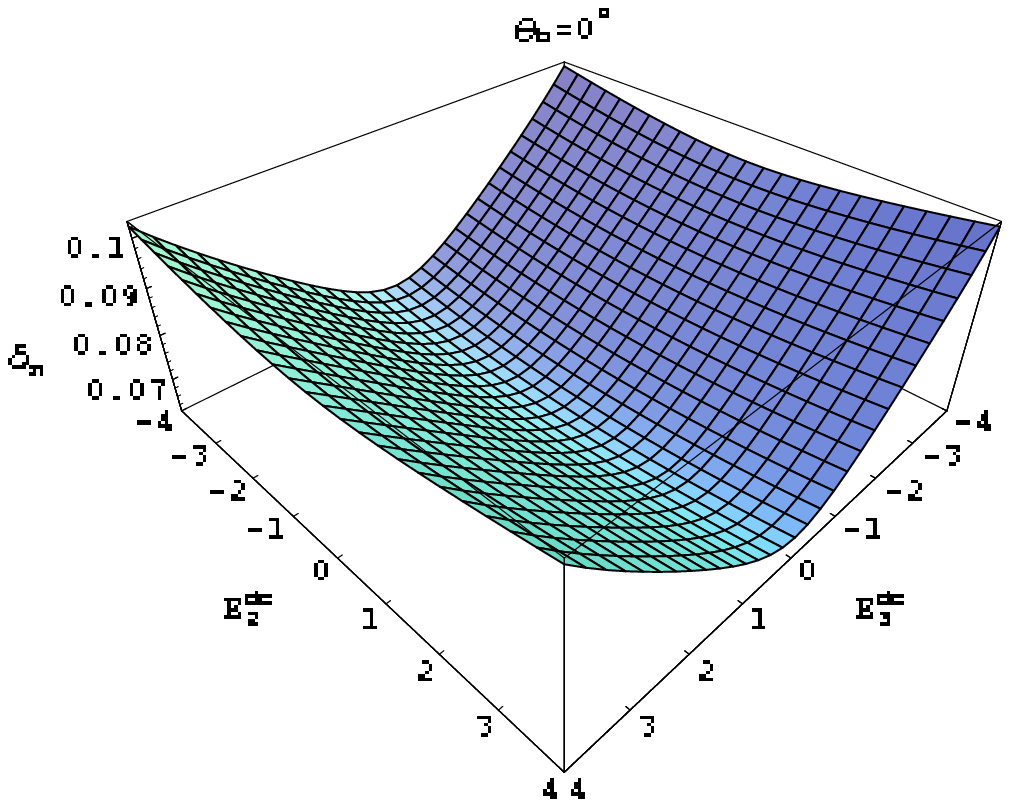,width=2.8in}\\
\vspace{10mm} \psfull\epsfig{file=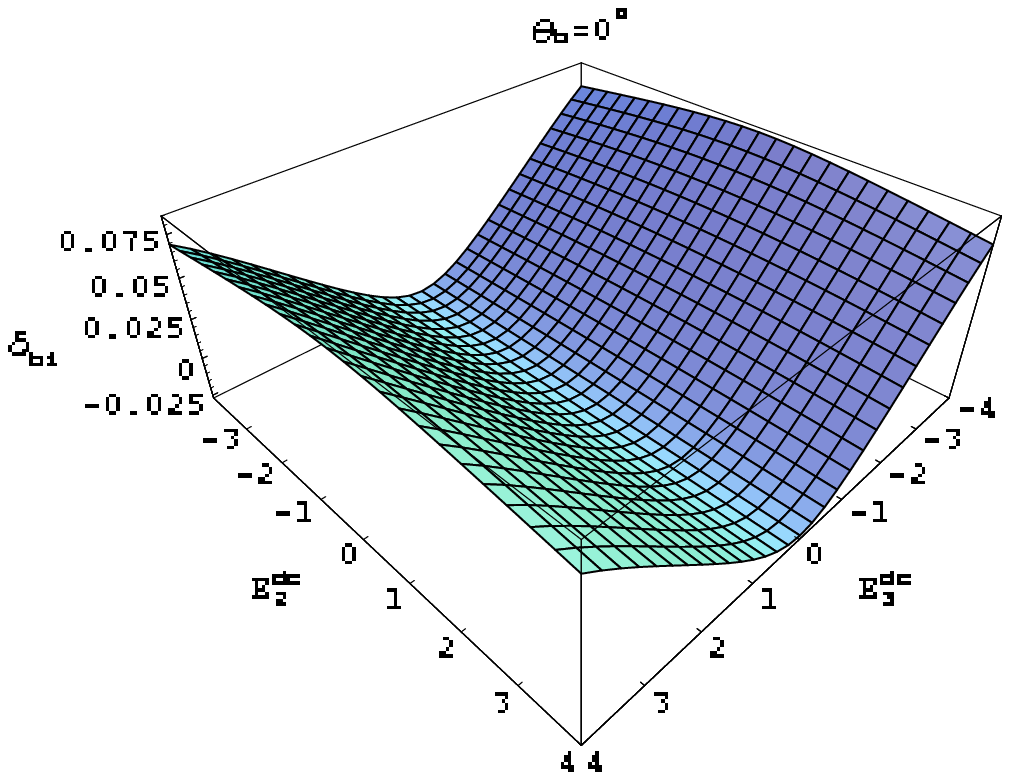,width=2.8in}
\hspace{10mm} \epsfig{file= 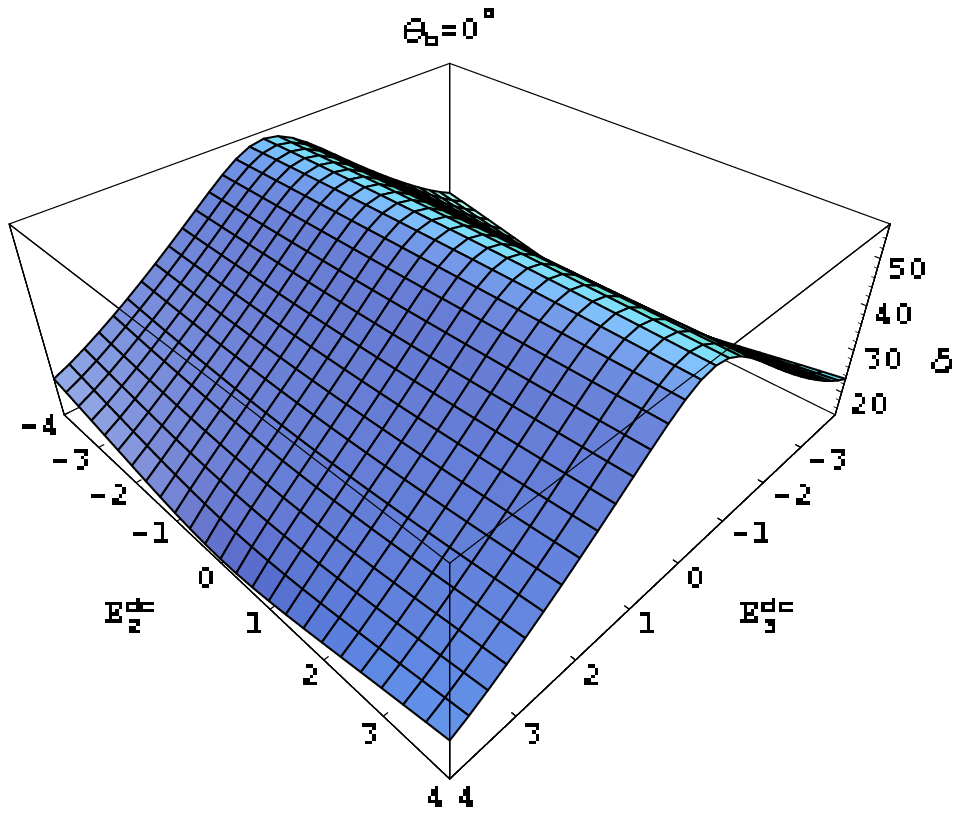,width=2.8in}
\\ \vspace{10mm}
\psfull\epsfig{file=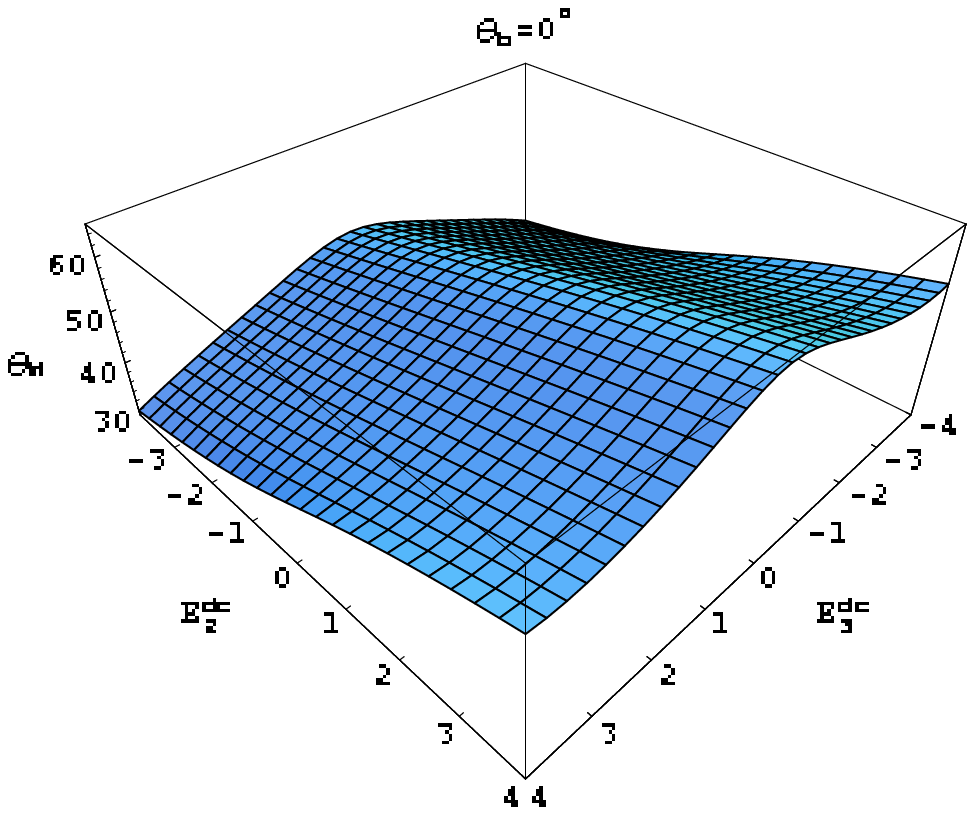,width=2.8in} \hspace{10mm}
\epsfig{file= 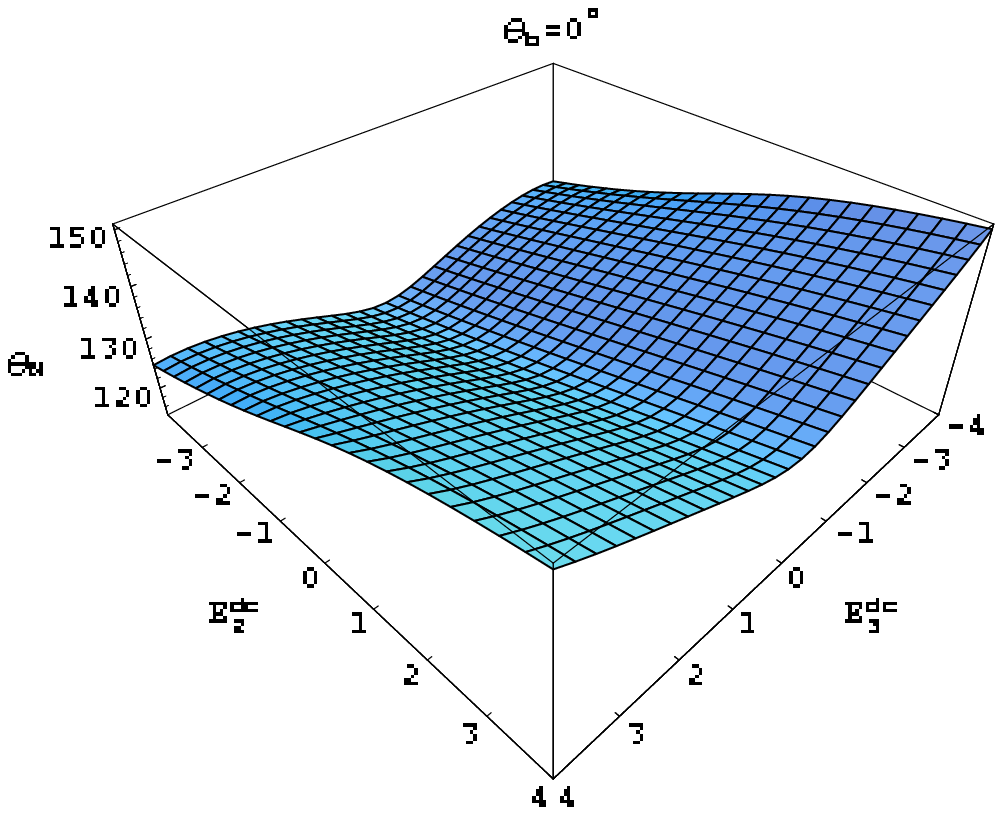,width=2.8in}
  \caption{\label{fig4} The HCM parameters $\eps^{Br}_2$,
  $\delta_n$, $\delta_{bi}$, $\delta$ (in degree) and $\theta_{M,N}$
  (in degree) plotted against $E^{dc}_{2,3}$ (in V m${}^{-1}$ $\times
  10^7$).
The crystallographic orientation angles
  of material $b$ are $\theta_b
= \phi_b = 0$; and $E^{dc}_1 = 0$. Material $b$ is potassium niobate, and the
particles of both constituent materials are spherical.}
\end{figure}

The HCM parameters are plotted in Figure~\ref{fig4} as functions of
$E^{dc}_{2,3}$  with $E^{dc}_1 =0$. As in Figure~\ref{fig2}, the crystallographic
orientation angles of
material $b$   are taken as $\theta_b = \phi_b =
0$.  Whereas the parameters $\eps^{Br}_2$, $\delta_n$,
$\delta_{bi}$, and $\delta$ are not particularly sensitive to
$E^{dc}_{2}$, they do vary significantly as $E^{dc}_{3}$ varies. Most notably,
the HCM can be made either negatively biaxial ($\delta_{bi}<0$) or
positively biaxial ($\delta_{bi}>0$).
The two optic axes of the HCM need not be mutually orthogonal,
with the included angle $2\delta$ between them as low as $40^\circ$.
The
polar angles $\theta_{M,N}$  are
sensitive to both $E^{dc}_{2}$ and $E^{dc}_{3}$. We note that the
sign of $E^{dc}_3$ does not influence any of the HCM parameters, but
the sign of $E^{dc}_2$ does influence the polar angles
$\theta_{M,N}$.

\begin{figure}[!ht]
\centering \psfull\epsfig{file=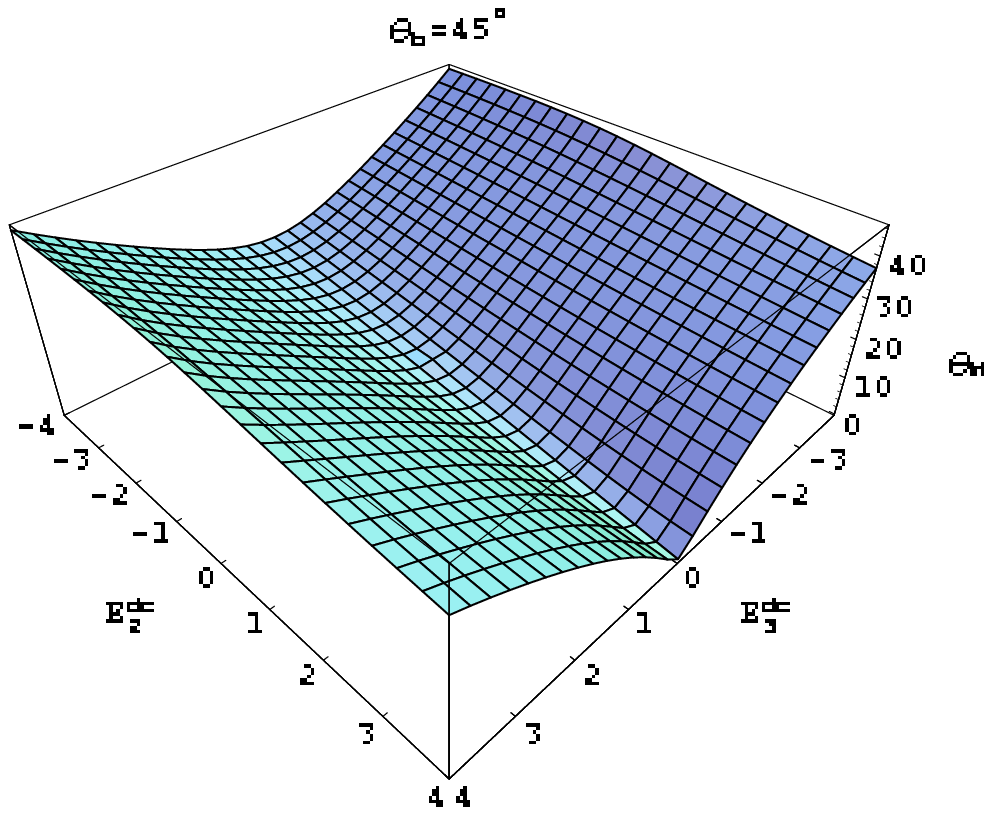,width=2.8in}
\hspace{10mm} \epsfig{file= 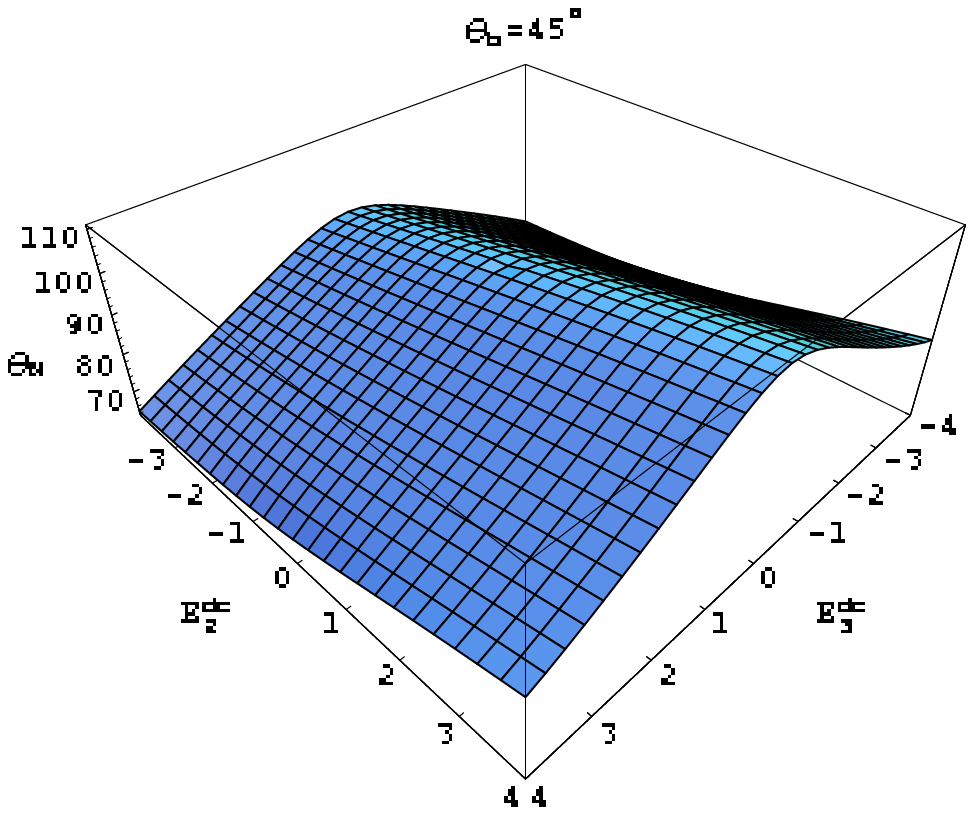,width=2.8in}
\\ \vspace{10mm}
\psfull\epsfig{file=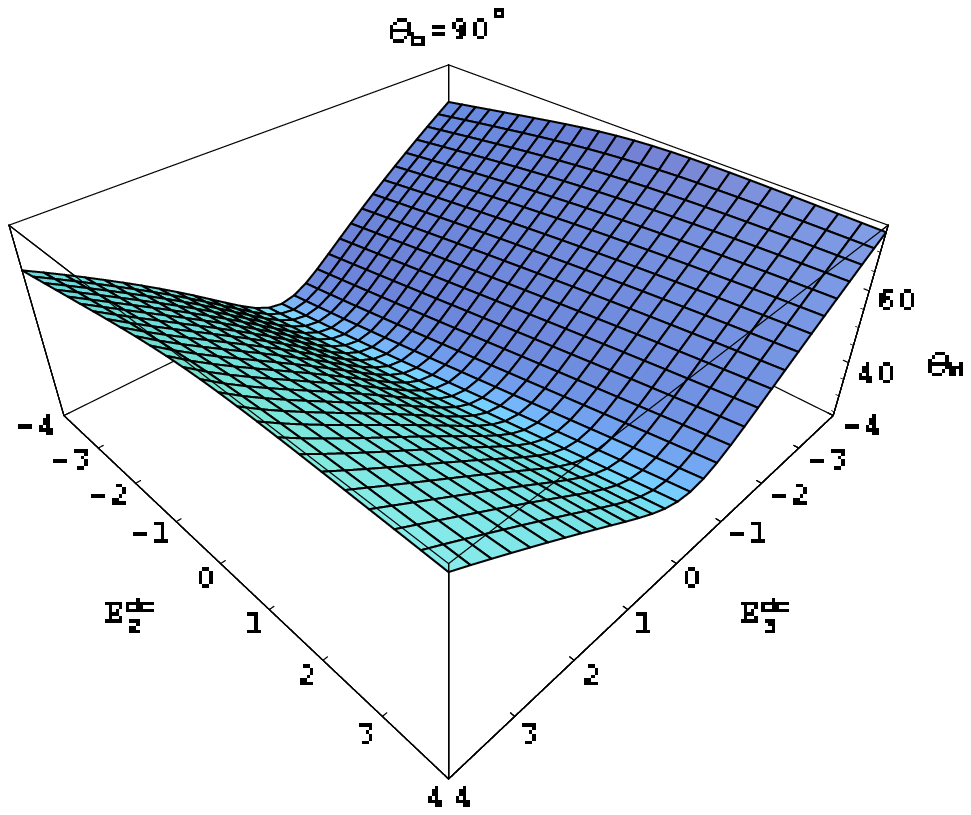,width=2.8in} \hspace{10mm}
\epsfig{file= 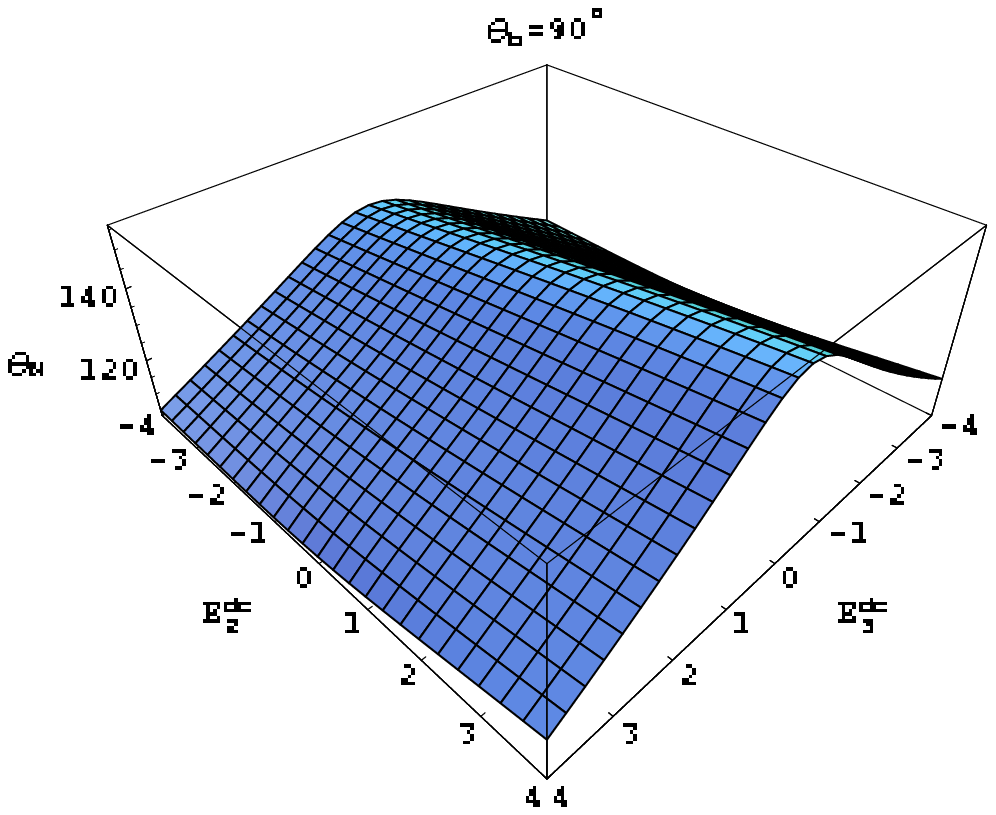,width=2.8in}
  \caption{\label{fig5}
The optic ray axis angles $\theta_{M,N}$  (in degree) plotted
against $E^{dc}_{2,3}$ (in V m${}^{-1}$ $\times
  10^7$). The constitutive parameters
  of material $b$ are same as in Figure~\ref{fig4}
except that $\theta_b \in \lec  45^\circ, 90^\circ \ric$. (The
corresponding plots of $\eps^{Br}_2$,
  $\delta_n$, $\delta_{bi}$,
  and $\delta$ are not noticeably different to  those presented in
Figure~\ref{fig4}.)}
\end{figure}

The influence of the  orientation angle $\theta_b$   is explored in
Figure~\ref{fig5}. The constitutive parameters of material $b$ are
the same as in Figure~\ref{fig4} but with  $\theta_b \in \lec
45^\circ, 90^\circ \ric $. As is the case in Figure~\ref{fig3}, the
graphs in Figure~\ref{fig5} show that the dependencies of the polar
angles $\theta_{M,N}$ upon the components of $\#E^{dc}$ are acutely
sensitive to $\theta_b$.
 The HCM parameters
$\eps^{Br}_2$, $\delta_n$, $\delta_{bi}$, and $\delta$~---~which are
not presented in Figure~\ref{fig5}~---~are insensitive to increasing
$\theta_b$; the plots for these quantities are not noticeably
different to the corresponding plots presented in Figure~\ref{fig4}.

A comparison of Figures \ref{fig2} and \ref{fig3} with Figures \ref{fig4}
and \ref{fig5} shows that the application of $\#E^{dc}$ is more effective
when material $b$ is potassium niobate rather than zinc telluride. A dc electric
field that is two orders smaller in magnitude is required for changing the HCM properties
with potassium niobate than with zinc telluride, and this observation is
reaffirmed by comparing the upper and lower graphs in Figure~\ref{fig1}.
To a great extent, this is
due to the larger electro--optic coefficients of potassium niobate; however,
we cannot rule out some effect of the crystallographic
structure of material $b$, which we plan to explore in the near future.

 Electrical control   appears to require dc electric fields of high magnitude.
 However, the needed dc voltages can be comparable
with the half--wave voltages of electro--optic materials (Yariv \& Yeh 2007). We
must also note that the required magnitudes of
$\#E^{dc}$ are much smaller than
the characteristic atomic electric field strength (Boyd 1992).
 The
possibility of electric breakdown  exists, but it would significantly
depend on the time that the dc voltage would be switched on for.
Finally, the non--electro--optic constituent
material may have to be a polymer that can withstand
high dc electric fields.

\subsection{Identically oriented spheroidal electro--optic particles}
\label{oriented_spheroids}

We close by considering the scenario wherein the effect of the Pockels effect is
going to be highly
noticeable in the HCM~---~that is, when the particles of material $b$
are highly aspherical and the crystallographic orientation as well as the
geometric orientation of these particles are
 aligned with $\#E^{dc}$. We chose potassium
niobate~---~which is more sensitive to the application of $\#E^{dc}$ than zinc
telluride~---~for our illustrative results.

\begin{figure}[!ht]
\centering \psfull\epsfig{file=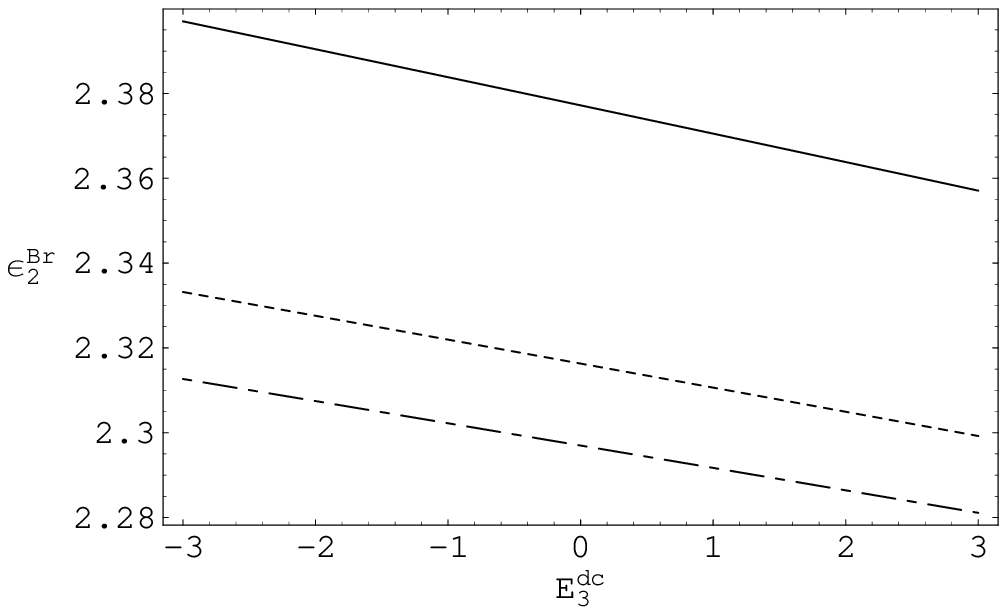,width=2.8in}
\hspace{10mm} \epsfig{file= 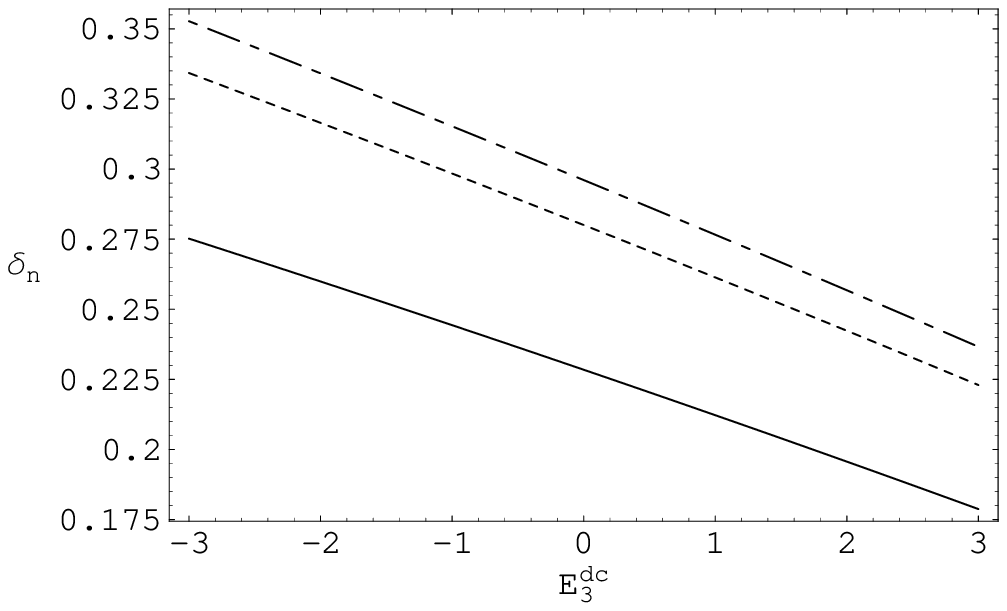,width=2.8in}\\
\vspace{10mm} \psfull\epsfig{file=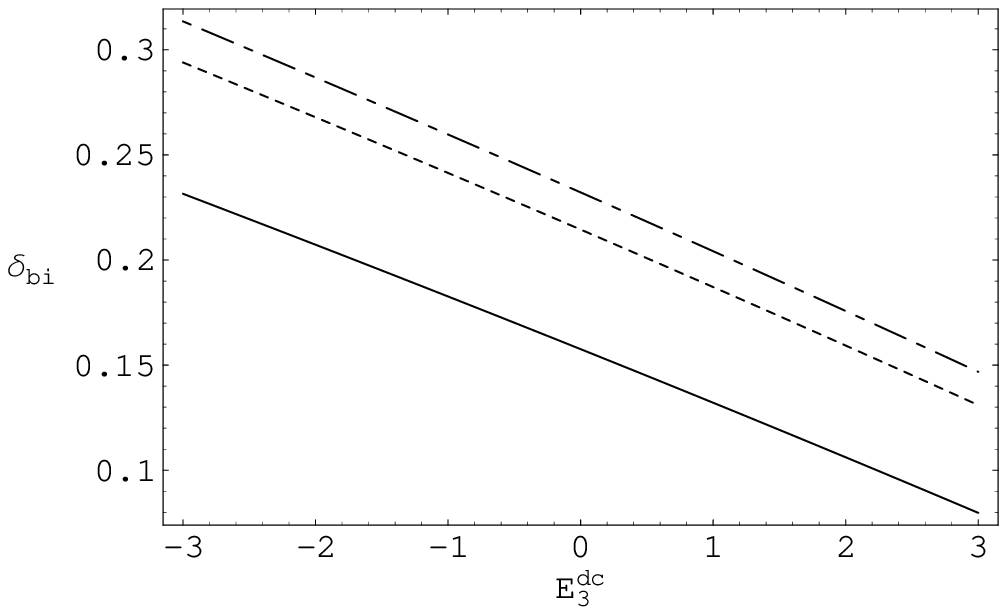,width=2.8in}
\hspace{10mm} \epsfig{file= 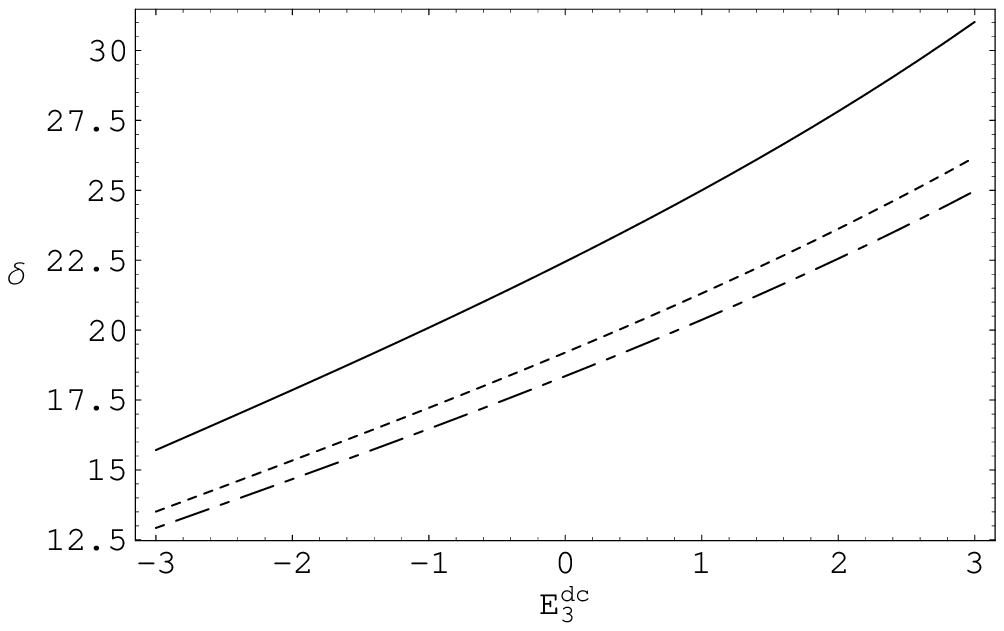,width=2.8in}
\\ \vspace{10mm}
\psfull\epsfig{file=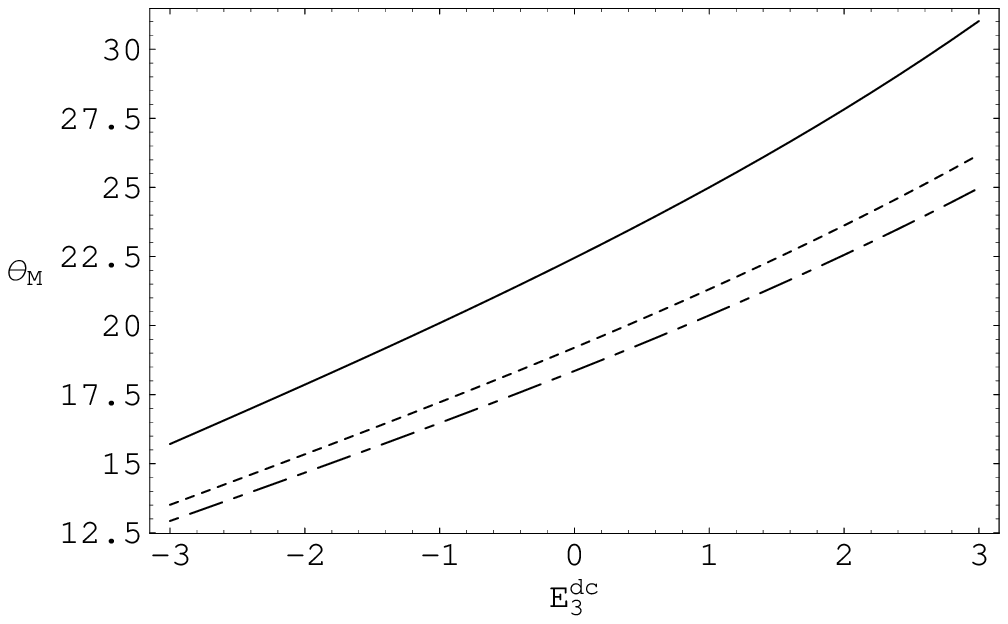,width=2.8in}
  \caption{\label{fig6} The HCM scalar parameters $\eps^{Br}_2$,
  $\delta_n$, and $\delta_{bi}$,
 $\delta$ (in degree), and $\theta_{M}$
  (in degree) plotted against $E^{dc}_{3}$ (in V m${}^{-1}$ $\times
  10^8$). The crystallographic orientation angles $\theta_b
= \phi_b = 0$, and $E^{dc}_{1,2} = 0$. Material $b$ is potassium
niobate. The constituent materials are distributed as spheroids with
shape parameters: $\alpha_1 = \alpha_2 = \beta_1 = \beta_2 = 1$; and
$\alpha_3 = \beta_3 = 3$ (solid curves), $\alpha_3 = \beta_3 = 6$
(dashed curves), and $\alpha_3 = \beta_3 = 9$ (broken dashed
curves). The optic ray angle $\theta_N = 180^\circ - \theta_M $.}
\end{figure}

In Figure~\ref{fig6}, the HCM parameters $\eps^{Br}_2$, $\delta_n$,
$\delta_{bi}$, $\delta$, and $\theta_{M}$ are plotted against
$E^{dc}_3$, with $E^{dc}_{1,2} = 0$. Both constituent materials are
distributed as identical spheroids with shape parameters $\alpha_{1,2} =
\beta_{1,2}= 1$ and $\alpha_3 = \beta_3 \in \lec  3, 6, 9 \ric$; furthermore,
$\theta_b=\phi_b=0$. As
$\theta_N = 180^\circ - \theta_M $ for this scenario,
$\theta_N$ is not plotted.    All
the presented HCM parameters vary considerably as $E^{dc}_3$
increases; furthermore, all are sensitive to the sign of $E^{dc}_3$.

We note that the degree of biaxiality and the linear birefringence
increase as $\alpha_3=\beta_3$ increases. This is a significant
conclusion because perovskites (such as potassium niobate) are
nowadays being deposited as oriented nanopillars (Gruverman \&
Kholkin 2006).

\section{Concluding remarks} \label{conc}

The homogenization of particulate composite materials
with constituent materials that can exhibit the
Pockels effect gives rise to HCMs  whose effective constitutive parameters may
be continuously varied through the application of a low--frequency (dc)
electric field. Observable effects can be achieved  even when the
constituent particles are randomly oriented. Greater
control over the HCM constitutive parameters may be achieved by
orienting the constituent particles. By homogenizing constituent
materials which comprise oriented elongated particles rather than
oriented spherical particles, the degree of electrical control over the HCM
constitutive parameters is further increased.  The vast panoply of complex
materials currently being investigated (Grimmeiss {\em et al.} 2002; Weiglhofer \& Lakhtakia 2003; Mackay
2005; Mackay \& Lakhtakia 2006) underscores the importance of electrically controlled composite materials
for a host of applications for telecommunications, sensing, and actuation.

\vspace{5mm}

 \noindent{\bf References}
\begin{enumerate}

\item
Auld, B.A. 1990 {\em Acoustic fields and waves in solids}.
Malabar, FL, USA: Krieger.

\item
Biot, J.--B. \&  Arago, F. 1806 M\'emoire sur les affinit\'es des corps pour la lumi\`ere
et parti\-culi\`erement sur les forces r\'efringentes des diff\'erents gaz.
\emph{M\'em. Inst. Fr.} {\bf 7}, 301--385.

\item
Bruggeman, D.A.G. 1935
Berechnung verschiedener physikalischer Konstanten von Substanzen. I. Dielektrizit\"atskonstanten und
Leitf\"ahgkeiten der Misch\-k\"orper aus isotropen Substanzen.
{\em Ann. Phys. Lpz.} {\bf 24}, 636--679. [Facsimile reproduced in Lakhtakia (1996).]

\item
Boyd, R.W. 1992 {\em Nonlinear optics}. San Diego, CA, USA: Academic Press.

\item
Bragg, W.L. \&   Pippard, A.B. 1953 The form birefringence of
macromolecules. {\em Acta Crystallogr.} {\bf 6}, 865--867.

\item
Chen, H.C. 1983 {\em Theory of electromagnetic waves: A coordinate--free
approach}. New York,
NY, USA: McGraw--Hill.

\item
Cook Jr., W.R. 1996 Electrooptic coefficients, in:  Nelson, D.F. (ed.),
{\em Landolt--Bornstein Volume III/30A}.
Berlin, Germany: Springer.

\item
Finkelmann, H., Kim, S. T., Mu\~noz, A., Palffy--Muhoray, P. \& Taheri, B. 2001
Tunable mirrorless lasing in cholesteric
liquid crystalline elastomers. {\em Adv. Mater.} {\bf 13}, 1069--1072.

\item
Gribble, C.D. \& Hall, A.J. 1992
{\em Optical mineralogy: Principles \& practice}. London,
United Kingdom: UCL Press.

\item
Grimmeiss, H.G., Marletta, G., Fuchs, H. \& Taga, Y. (eds.) 2002
{\em Current trends in nanotechnologies: From materials to systems}. Amsterdam,
The Netherlands: Elsevier.

\item
Gruverman, A. \& Kholkin, A. 2006
Nanoscale ferroelectrics: processing, characterization, and future
trends. {\em Rep. Prog. Phys.} {\bf 69}, 2443--2474.

\item
Klein, C. \& Hurlbut, Jr., C.S. 1985
{\em Manual of mineralogy}. New York, NY, USA: Wiley. (pp. 247 {\em et seq.})

\item
Lakhtakia, A. 1993
Frequency--dependent continuum properties of a gas of scattering
centers. {\em Adv. Chem. Phys.} {\bf 85}(2), 311--359.

\item
Lakhtakia, A. (ed.) 1996
{\em Selected papers on linear optical composite materials}. Bellingham, WA, USA: SPIE Optical
Engineering Press.

\item
Lakhtakia, A. 2006a
Electrically tunable, ultra\-narrow\-band, circular--polarization rejection filters
with electro--optic structurally chiral materials. {\em J. Eur. Opt. Soc.~--~Rapid Pubs.}
{\bf 1}, 06006.

\item
Lakhtakia, A. 2006b
Electrically switchable exhibition of circular Bragg phenomenon by an isotropic slab. {\em Microw. Opt. Technol. Lett.}
{\bf 48}, at press.

\item
Lakhtakia, A., McCall, M.W.,  Sherwin, J.A., Wu, Q.H. \&
Hodgkinson, I.J. 2001
Sculptured--thin--film spectral
holes for optical sensing of fluids.
{\em Opt. Commun.} {\bf  194}, 33--46.

\item
Mackay, T.G. 2005 Linear and nonlinear
homogenized composite mediums as metamaterials.
{\em Electromagnetics} {\bf 25}, 461--481.

\item
Mackay, T.G. \& Weiglhofer, W.S.  2001 Homogenization of biaxial
composite materials: nondissipative dielectric properties. {\em
Electromagnetics} {\bf 21}, 15--26.

\item
Mackay, T.G. \& Lakhtakia, A. 2006
Electromagnetic fields in linear bianisotropic mediums. {\em Prog. Opt.}
(at press).

\item
Michel, B. 2000,
Recent developments in the homogenization
of linear bianisotropic composite materials. In:
Singh, O.N. \&  Lakhtakia, A. 2000
{\em Electromagnetic fields in unconventional materials
and structures}.
New York, NY, USA: Wiley.

\item
M\"onch, W., Dehnert, J., Prucker, O., R\"uhe, J. \& Zappe, H. 2006
Tunable Bragg filters based on polymer swelling.
{\em Appl. Opt.} {\bf 45}, 4284--4290.

\item
Neelakanta, P.S. 1995
{\em Handbook of electromagnetic materials~---~Monolithic and composite versions
and their applications}. Boca Raton, FL, USA: CRC Press.

\item
Schadt, M. \& F\"unfschilling, J. 1990
New liquid crystal polarized color projection principle.
{\em Jap. J. Appl. Phys.} {\bf 29}, 1974--1984.

\item
 Shafarman, W.N.,  Castner, T.G.,
Brooks, J.S.,   Martin, K.P. \&  Naughton, M.J. 1986
Magnetic tuning of the metal--insulator transition for uncompensated arsenic--doped silicon.
{\em Phys. Rev. Lett.} {\bf 56}, 980--983.

\item
Sherwin, J.A.  \&   Lakhtakia, A. 2002
Bragg--Pippard formalism for bianisotropic particulate composites.
{\em Microw. Opt. Technol. Lett.} {\bf 33}, 40--44.

\item
Yariv, A.  \& Yeh, P. 2007 {\em Photonics: Optical electronics in modern
communications, 6th ed}.
New York, NY, USA: Oxford University Press.

\item
Yu, H.,   Tang, B.Y.,    Li, J. \& Li, L. 2005
Electrically tunable lasers made from electro--optically active photonics band gap materials.
{\em Opt. Express} {\bf 13}, 7243--7249.

\item
Wang, F., Lakhtakia, A. \& Messier, R. 2003
On piezoelectric control of the optical response of sculptured thin films.
\emph{J. Modern Opt.} {\bf 50}, 239--249.

\item
Weiglhofer, W.S. \& Lakhtakia, A. 1999 On electromagnetic waves in
biaxial bianisotropic media. {\em Electromagnetics} {\bf 19},
351--362.

\item
Weiglhofer, W.S. \& Lakhtakia, A. (ed.) 2003 {\em Introduction to
complex mediums for optics and electromagnetics}. Bellingham, WA,
USA: SPIE Press.

\item
Weiglhofer, W.S., Lakhtakia, A. \& Michel, B. 1997
Maxwell Garnett and Bruggeman formalisms for a particulate
composite with bianisotropic host medium.
{\em Microw. Opt. Technol. Lett.} {\bf 15}, 263--266; correction:
1999 {\bf 22}, 221.

\end{enumerate}

\newpage

\newpage

\newpage

\newpage

\newpage

\newpage

\end{document}